\documentclass[12pt]{article}

\usepackage{epsf}
\usepackage{amsmath}
\usepackage{graphics}
\usepackage{cite}
\usepackage{color}

\renewcommand{\thefootnote}{\#\arabic{footnote}}
\renewcommand{\thefootnote}{\fnsymbol{footnote}}
\begin{document}

\def\thefootnote{\fnsymbol{footnote}}

\begin{center}
\Large
{\bf Memories of Steven Weinberg (1933-2021)} \\ 
\normalsize
\vspace{1.0in}

Paul H. Frampton

Dipartimento di Fisica, Universit`a del Salento
and INFN Sezione di Lecce, \\Via Arnesano 73100 Lecce, Italy.

\vspace{1.0in}

{\bf Abstract}
\end{center}
Steven Weinberg, renowned particle theorist and Nobel laureate, passed away in July 2021. We discuss selections
of his work on
effective field theory, electroweak unification, and symmetry related topics. We then add a few memories of Weinberg at Harvard University then at the University of Texas, Austin.

\section{Introduction}

\noindent
 Weinberg's legacy in theoretical physics is embodied within 300 papers and 17 books,
 of which 8 were technical and 9 were popular.
 According to his own writings, he regarded his two greatest accomplishments in rank
 order as (i) effective field theory; (ii) electroweak unification. His Nobel prize was awarded
 for (ii), not (i), but a similar complaint, if it is a complaint, could equally be made by another
 renowned particle theorist, C.N. Yang. 
 
 \bigskip
 
 \noindent
 \underline{His CV:}\\
 Bronx High School of Science, graduated 1950 (as did Glashow).\\ 
 Bachelor degree Cornell 1954; PhD Princeton (Treiman) 1957.\\ 
 Postdoc: Columbia 1957-59. \\
 Faculty: Berkeley 1959-66; MIT 1966-73; \\
 Harvard 1973-82; UT-Austin 1982-2021.\\ 

\noindent
In our opinion, his best books were :\\
Gravitation and Cosmology 1972 (technical)\\
The First Three Minutes 1977 (popular)\\ 
Both books remain fresh. \\

 \noindent
 Steven Weinberg died on July 23, 2021 in Austin, Texas,\\
 the same day as Toshihide Maskawa in Kyoto, Japan.\\
 
\noindent
We heard Weinberg give a few talks in the early 1970s before meeting him
personally when we were a postdoc at Harvard for the 3 years 1978-80.
In 1982 he left Harvard to Texas and invited
us to spend the year 1982-83
with his group at  UT-Austin. We were therefore his colleague for a total of
four years.\\

\noindent
The adjectives about him which come to mind are he was intense and competitive. 
He was not given to small talk and was always serious.\\

\noindent
\section{Effective field theory}

\noindent
Weinberg's most impactful work was provided by his monumental papers
{\it e.g.} \cite{Weinberg1979,Weinberg1980,Weinberg1990,Weinberg1991,Weinberg1996},
spread over more than a decade and a half, which developed the techniques
and formalism of effective field theory. There is a lot of scientific honesty in the fact that 
although his standard model of electroweak
unification (see below) was discovered in part by imposing strict renormalisability, 
Weinberg then spent far more time
examining nonrenormalisable theories and showing convincingly how in many cases
they can be comparably as predictive.

\bigskip

\noindent
A simple effective field theory appears when we observe that when we add to a renormalisable
lagrangian ${\cal L}_{ren}$ which contains only terms of dimension $d=4$, higher-dimension
terms with $d=5,6,....$ the full lagrangian can be expected, under appropriate assumptions, to
appear in the form
\begin{equation}
{\cal L} = {\cal L}_{ren} + \Sigma_i  \frac{C_i^{(5)}}{\Lambda} {\cal L}_i^{(5)} +\Sigma_i \frac{ C_i^{(6)} }{\Lambda^2}{\cal L}_i^{(6)} +
...
\label{calL}
\end{equation}
where $\Lambda$ is a cut-off scale much higher then the energies typically appearing in the
theory described by ${\cal L}_{ren}$ alone.

\bigskip

\noindent
For example, let us consider that the starting renormalisable lagrangian is the standard model
${\cal L}_{ren} \equiv {\cal L}_{SM}$. In perturbation theory the standard model conserves
lepton number $L$ and baryon number $B$. $L$ violation first appears in $d=5$ terms which can generate
Majorana neutrino masses so an appropriate cut-off could be a see-saw mass $\lambda \sim 10^{10}$ GeV.
$B$ violation first appears  in $d=6$ terms so there an appropriate cut-off could be a grand unification
scale $\Lambda \sim 10^{15}$ GeV. In both cases, the contributions will be suppressed by
\begin{equation}
\left( \frac{E}{\Lambda} \right)^{(d-4)}
\label{suppress}
\end{equation}
where $E$ is a characteristic energy scale of the process in question.

\bigskip

\noindent
Weinberg's starting point \cite{Weinberg1979} was to show how current algebra results first obtained
in the 1960s can be derived without the use of current algebra by the use of phenomenological
lagrangrians which satisfy the assumed symmetry principles and studying Feynman diagrams.
Underlying this success is the fact that the most general quantum field theory has essentially no content
except analyticity, unitarity, cluster decomposition and symmetry.

\bigskip

\noindent
This procedure allows the calculation of corrections to the current algebra results which
were found much earlier in the 1960s including by Weinberg himself. Long before effective
field theory, our own modest 1968 Oxford DPhil thesis under John Taylor showed how certain
current algebra sum rules could be derived without current algebra by using analyticity
in the form of superconvergence sum rules. In his next paper\cite{Weinberg1980}, Weinberg applied
the effective field theory technique to gauge theories of grand unification as briefly
mentioned {\it ut supra}.

\bigskip

\noindent
In the early 1990s buoyed by his successes with effective field theory
in particle theory, Weinberg extended the use of effective field theory to 
nuclear physics. In 1990 \cite{Weinberg1990},
he calculated the forces between two or more nucleons from chiral symmetry breaking.
Then in 1991\cite{Weinberg1991} he extended this to include an arbitrary number
of soft pions.

\bigskip

\noindent
In impressive formal work \cite{Weinberg1996} in 1996, Weinberg discussed when a gauge theory which is nonrenormalisable
in the usual power-counting sense can be nevertheless renormalisable in the sense that
all divergences can be cancelled by a renormalisation of the infinite number of terms in the bare action.
It was shown in \cite{Weinberg1996} that this is true if the constraint on the bare action
corresponds to the cohomology of the BRST transformations generated by the action.

\bigskip

\noindent
The last paper \cite{Weinberg2021} archived by Weinberg, only six months before his death,
was a review of effective field theory. This supports the idea that he believed it to be his greatest
legacy. Paper \cite{Weinberg2021} is recommended to the reader as a
beautifully-written retrospective.

\bigskip

\section{Electroweak unification}

\noindent
To discover his electroweak unification\cite{Weinberg1967}, Weinberg adopted
the $SU(2) \times U(1)$ gauge group proposed by Glashow several years ealier and
combined that with the Higgs, or better Brout-Englert-Higgs, mechanism for
spontaneous symmetry breaking. His specific contribution was to introduce the
scalar Higgs doublet
\begin{equation}
H = \left( \begin{array}{c}  H^+ \\ H^0 
\end{array} \right)
\label{Higgsdoublet}
\end{equation}
and to allow the vacuum expectation value $<H^0>$ of $H^0$ to induce spontaneous symmetry
breaking of $SU(2) \times U(1)$ into the $U(1)_{QED}$ of quantum electrodynamics.
The resultant theory contains three massive gauge bosons $W^{\pm},  Z^0$, together with
the massless photon of QED and one physical scalar field, the Higgs Boson. While Weinberg 
did not introduce any new particle because $Z^0$ and the Higgs Boson had already been successfully
predicted, his choice of the dimension $d=2$ doublet in Eq.(\ref{Higgsdoublet}) was important.
He could have added $SU(2)$ representations of higher dimensions $d=3, 4, 5, ....$ but
the choice in Eq.(\ref{Higgsdoublet}) as a central part of the standard model has led to agreement
with every experiment for the last six decades. This alone provides Weinberg with multi-century legacy.
The choice allowed a prediction of the $M(W)/M(Z)$ mass ratio in terms of the electroweak
mixing angle, in agreement with subsequent experiments.

\bigskip

\noindent
At the time in 1967 weak interactions were not at the forefront of particle theory research.
We were a DPhil student in Oxford at the time when Weinberg published no less than six interesting
papers in Physics Review Letters and we were recommended to read five of them,
but never \cite{Weinberg1967}.  More generally paper\cite{Weinberg1967}
was hardly cited at all until 1971. In 1971, a young Gerard 't Hooft created a sensation by
demonstrating that such a theory is
renormalisable on a footing with QED.
At the end of \cite{Weinberg1967}, Weinberg had conjectured that renormalisability might
survive spontaneous symmetry breaking but could not prove it.

\bigskip

\noindent
As soon as 't Hooft's proof appeared in 1971, Weinberg\cite{Weinberg1971} revisited
the theory and showed how all the infinities cancelled in processes such as
\begin{eqnarray}
\nu + \nu & \rightarrow & W^+ + W^- \nonumber \\
\nu + \nu & \rightarrow & \nu + \nu. \nonumber \\
e^+ + e^- & \rightarrow & W^+ + W^- \nonumber\\
\label{processes}
\end{eqnarray}
and remarked that the experimental discovery of the $W^{\pm}, Z^0$ gauge bosons
would provide the best confirmation of the theory, as transpired in the 1980s.

\bigskip

\noindent
Depsite its slow start, paper\cite{Weinberg1967} is now one of the two most cited
papers ever in particle theory. It is the paper for which Weinberg shared the 1979
Nobel prize with Glashow and Salam.

\section{Symmetry Related Topics}

\noindent
Weinberg published deep papers about symmetry which plays a central
r\"ole in particle theory. Of these we shall discuss four 
\cite{SWeinberg1962,SWeinberg1974,SWeinberg1976,SWeinberg1979} 
which were particularly impactful.

\bigskip

\noindent
In 1962, one of the hottest topics was the Nambu-Goldstone theorem
that spontaneous breaking of a global symmetry leads to the existence
of massless spin-zero states. Nambu had reached this conclusion in 1960 from
his deep understanding of the BCS theory of superconductivity, and went 
on a few years later to write related papers with Jona-Lasinio.  Several months after
Nambu, Goldstone 
provided a mathematical model which showed clearly how the N-G bosons
arise for a Mexican-hat potential. Goldstone modestly claimed that
his mathematical model had no physical application despite citing Nambu
who had actually provided one: the lightness of the pion is a result of the
spontaneous breaking of chiral symmetry. For decades the 
massless or light states were called Goldstone bosons and only recently
Nambu-Goldstone bosons. Fairest might be Nambu bosons. The problem 
for Nambu was a partially inscrutable writing style, not as perspicuous as 
that of Goldstone or indeed Weinberg. Nevertheless, the physics is there 
in Nambu. Despite this digression, we shall call it the N-G Theorem,

\bigskip

\noindent
In \cite{SWeinberg1962}, with Goldstone and Salam, Weinberg made a
broad discussion of spontaneous breaking of global symmetries in
quantum field theory with a view to proving the N-G theorem in the
most general way possible. Indeed GSW succeeded to provide proofs, first within
perturbation theory and then more generally. GSW\cite{SWeinberg1962}
thus convinced the community in 1962 that for global symmetries the N-G
theorem is correct. With the benefit of hindsight, we now know that
when the global symmetry is replaced by a local gauged symmetry
the N-G theorem no longer holds,
as shown by Brout, Englert and Higgs in 1964.
Nevertheless GSW was a timely and pivotal contribution to the emerging 
theory.

\bigskip

\noindent
We proceed in chronological order to Weinberg's 1974 solo paper
(the majority of his papers were solo) which discussed\cite{SWeinberg1974} spontaneously
broken global and local symmetries at high temperatures. In the majority
of cases (there are a few exceptions) the symmetry will be restored at
a critical temperature $T=T_C$ at which there is a phase transition.
Weinberg considered a globally $O(N)$ symmetric scalar field theory with lagrangian
\begin{equation}
{\cal L} = -\frac{1}{2} \partial_{\mu}\phi_i \partial^{\mu} \phi_i - P(\phi)
\end{equation}
with a quartic polynomial potential
\begin{equation}
P(\phi) = \frac{1}{2} {\cal M}_0^2 \phi_i\phi_i + \frac{1}{4} e^2 (\phi_i\phi_i)^2
\end{equation}
in which $\phi_i$ is an n-dimensional vector representation of $O(n)$ and
he is considering the spontaneous breaking $O(n) \rightarrow O(n-1)$.

\bigskip

\noindent
The calculation at high temperature $T$ reveals that the potential becomes
\begin{equation}
P_{eff}(\phi) = \frac{1}{2} {\cal M}^2 (T) \phi_i\phi_i + \frac{1}{4} e^2(\phi_i\phi_i)^2
\end{equation}
in which
\begin{equation}
{\cal M}^2 (T) = {\cal M}^2 (0) + \frac{1}{12} (n+2) e^2 T^2
\end{equation}
The critical temperature $T_C$ is that for which
\begin{equation}
{\cal M}^2 (T_C) = 0
\end{equation}
which gives
\begin{equation}
T_C = \left( \frac{12}{n+2} \right)^{\frac{1}{2}} \frac{|{\cal M}(0)|}{e}
\end{equation}
The result of the spontaneous breaking $O(n) \rightarrow O(n-1)$ is 
to produce $(n-1)$ massless N-G bosons and one massive scalar of mass $M(0)$.
As usual, in the Mexican hat potential
\begin{equation}
m^2(T) = -2 {\cal M}^2(T)
\end{equation}
so that the critical temperature can finally be expressed as
\begin{equation}
T_C = \left( \frac{6}{n+2} \right)^{\frac{1}{2}} \frac{M(0)}{e}
\end{equation}
Weinberg proceeded in \cite{SWeinberg1974} to analyse more examples including local $O(n)$
and global $O(n) \times O(n)$.

\bigskip

\noindent
Although in the particle theory of $e^+e^-$. $e^-p$ and $pp$ colliders, we generally deal with $T=0$,
high $T$ becomes relevant in heavy-ion collisions and in the early universe
where phase transitions play a central r\"ole.

\bigskip

\noindent
One of the perhaps questionable features of the standard model was the existence
of elementary scalars. In 1976, Weinberg \cite{SWeinberg1976} attempted
to invent a theory with no spin-zero fields and no bare fermion masses.
With $N$ fermion types, not including colour,  such a theory possesses a
$U(N) \times U(N)$ global symmetry which is broken intrinsically by
electroweak interactions and spontaneously by strong interactions,
 the latter giving rise to pseudo-NG bosons.
Such a theory mandates the existence of extra-strong interactions, now
called technicolor, whose dynamics can be modeled by scaling QCD
to an energy scale $\Lambda_{TC}$, higher than $\Lambda_{QCD}$.
Weinberg\cite{SWeinberg1976} discussed some of the problems with making
a realistic theory.
Technicolor model builders have introduced extended technicolor
which subsumes technicolor into a larger gauge group. Despite
many attempts, it is fair to say that no realistic
technicolor theory is presently known.

\bigskip

\noindent
As a final example of Weinberg's work on symmetries, we discuss his
1979 paper\cite{SWeinberg1979} on the breaking of lepton (L)
and baryon (B) number symmetries by higher dimensional operators
in the standard model. L is first violated at dimension d=5. B is
first violated at d=6. Weinberg classified the d=6, $|\Delta B| \neq 0$ 
operators in the standard model using the notation for quarks
\begin{equation}
q_{i \alpha a L} ~~~~~~ i=1,2~~~ SU(2)_L;  ~~~~~\alpha=1,2,3  ~~ (colour); a=1,2,3 ~~~ (families).\\
\end{equation}
\noindent
and
\begin{equation}
u_{\alpha a R}; ~~~ d_{\alpha a R}
\end{equation}
\noindent
while, for leptons, he employed
\begin{equation}
l_{iaL} ~~~~~~~ i=1,2 ~~~ (SU(2)_L); ~~~~~~ a=1,2,3 ~~~ (families)
\end{equation}
\noindent
and
\begin{equation}
l_{a R}.
\end{equation}

\bigskip

\noindent
There are six independent d=6 B-violating operators, classified in \cite{SWeinberg1979} as follows\\
\begin{eqnarray}
{\cal O}_{abcd}^{(1)} & = & (\bar{d}^c_{\alpha a R} u_{\beta b R})(\bar{q}^c_{i \gamma c L}l_{jdL}) \epsilon_{\alpha\beta\gamma} \epsilon_{ij}, \nonumber \\
{\cal O}_{abcd}^{(2)} & = & (\bar{q}^c_{i\alpha a L}q_{j \beta b L})(\bar{u}^c_{\gamma c R}l_{dR}) \epsilon_{\alpha\beta\gamma} \epsilon_{ij},\nonumber \\
{\cal O}_{abcd}^{(3)} & = & (\bar{q}^c_{i \alpha a L} q_{j \beta b L}) (\bar{q}^c_{k \gamma c L} l_{ i d L}) \epsilon_{\alpha\beta\gamma} \epsilon_{ij} \epsilon_{kl},\nonumber \\
{\cal O}_{abcd}^{(4)} & = & (\bar{q}^c_{i \alpha a L} q_{j \beta b L}) (\bar{q}^c_{k \gamma c L} l_{i d L})\epsilon_{\alpha\beta\gamma} (\underline{\tau} \epsilon)_{ij} . 
(\underline{\tau} \epsilon)_{kl}  ,\nonumber \\
{\cal O}_{abcd}^{(5)} & = & (\bar{d}^c_{\alpha a R} u_{\beta b R}) (\bar{u}^c_{\gamma c R} l_{dR}) \epsilon_{\alpha\beta\gamma} ,\nonumber \\
{\cal O}_{abcd}^{(6)} & = & (\bar{u}^c_{\alpha a R} u_{\beta b R}) (\bar{d}^c_{\gamma c R} l_{dR}) \epsilon_{\alpha\beta\gamma} . \nonumber \\
\label{d6}
\end{eqnarray}

\bigskip

\noindent
The selection rule that $(B-L)$ is conserved follows from Eq. (\ref{d6}), as do some other constraints
which were discussed in \cite{SWeinberg1979}.

\bigskip

\noindent
Weinberg also considered d=5 terms in the standard model:
\begin{equation}
f_{abmn}\bar{l}^c_{i a L} l_{jbL}\phi_k^{(m)}\phi_l^{(n)}\epsilon_{ik}\epsilon_{jl}
+f_{abmn}^{'}\bar{l}^c_{1 a L}l_{jbL}\phi_k^{(m)}\phi_l^{(n)}\epsilon_{ij}\epsilon_{kl}
\end{equation}
These terms violate lepton number and can contribute non-zero Majorana neutrino masses.
We recall a discussion in 1979 at the Harvard faculty club when Weinberg said he had become
convinced that neutrinos have non-zero mass, 

\bigskip

\noindent
For proton decay, the fact that the d=6 operators in Eq.(\ref{d6}) respect a global
symmetry $(B-L)$ leads to the prediction that the leading proton decay is
\begin{equation}
p \rightarrow e^+ + \pi^0
\end{equation}
while the following decay should be suppressed
\begin{equation}
p \rightarrow e^- + \pi^+ +\pi^+
\end{equation}
If and when proton decay is observed, it will be interesting to confirm that the
second decay mode is essentially absent.

\bigskip

\noindent
We expect that Weinberg will be posthumously proven correct.

\noindent
\section{Harvard particle theory 1978-80}

\noindent
The faculty were Weinberg, Glashow, Coleman, Glauber. \\ 
\noindent 
More junior were Georgi, Witten, Carlson, Bagger ... \\ 
Visitors included Cahill, Sudarshan, Baulieu, Frampton ...\\

\noindent
This particle theory group was arguably
the best in the world.\\

\noindent
We sometimes occupied a desk in a corner of Sheldon Glashow's office and spent
mornings and afternoons talking to him and, in between, lunch at the faculty club with
Steven Weinberg.
To write particle theory papers, such a stimulating environment was perfect.\\

\noindent
Every Wednesday at noon was an unmissable gauge seminar.

\bigskip
\bigskip

\noindent
\underline{A couple of memories from Harvard:}\\

\noindent
(1) At one gauge seminar, Weinberg gave a talk about SusyGUTs, including saying
that new d=5 operators were harmless with respect to proton decay.
Georgi gently intervened with "That's an interesting phenomenological
question". Weinberg looked surprised, thought for a while (he was not too quick,
but deep) then said 
"Thank you Howard.
You have saved me embarrassment".
He had decided that the remark was correct. This was just one example
of Weinberg's scientific honesty.

\bigskip

\noindent
(2). At a faculty club lunch with Weinberg, the topic was: Does supersymmetry solve the gauge hierarchy problem?
Steve pointed out that even with supersymmetry the hierarchy remains an input and is definitely not an output.
Because of this, his conclusion was a categoric no.
We sometimes thought about Steve's conclusion during the next decade when paper after paper
stated the opposite. However, we never doubted that the correct answer is no.
Weinberg was an exceptionally good particle theorist.

\section{His move from Harvard to Austin}

\noindent
In the 1970s Louise Weinberg was teaching in a 3rd-rate college
in Boston. Steve said privately that if Louise got a job at a first-rate
place, he would move with her. In 1980, she received an offer from
the top-ten law school at UT-Austin.

\bigskip

\noindent
As a world-renowned physicist, Weinberg succeeded 
to negotiate a remarkable contract:

\begin{itemize}
\item
He never needed to retire from UT-Austin.
\item
During his lifetime, no other faculty member in the College of Arts and Sciences
at UT-Austin could ever earn more than 95\% of his salary.\\
(UT-Austin's president was allegedly close to tears on this one)
\item
Least importantly, unlimited credit at the UT-Austin faculty club.
\end{itemize}

\bigskip

\noindent 
Until the last day, Weinberg harboured doubts about leaving Harvard. We recall sitting
outside with Steve at a Harvard Square coffee establishment when he pulled from his
briefcase an unsigned letter resigning tenure which he studied assiduously.
It happened to be due that very day Wednesday, June 30th 1982. 
Steve said it was difficult to sign, the only time we ever saw Weinberg in doubt.\\

\noindent
The period 1980-82 provided Glashow with an opportunity to have some fun. 
Because of our relative youth,  we have no idea how they were in 1950 when graduating together
from the Bronx High School of Science but what we would say is that three decades 
later Shelly Glashow and Steve Weinberg had very different
personalities.
Shelly summarised the Weinberg situation by saying:
"An irresistible force has met an immovable object!"\\ 

\noindent
Glashow had a buddy at Texas A\&M, a different campus from UT-Austin
in the same state. He was an experimental particle physicist 
who had spent some time at Harvard,  greatly appreciated SG and smoothly arranged
a job offer to Glashow with double SW's new salary and equal 
to that of the Texas A\&M 
football coach. \\ 

\noindent
This accidental equality did not escape attention.
A journalist from the magazine Sports Illustrated asked Glashow
what he thought about being paid
the same as a football coach. Shelly's quick and fun reply was:
"At Harvard, I earn more than the football coach!"

\section{UT-Austin 1982-83}

\noindent
Weinberg was starting up his particle theory group in Austin. 
Each week there was a brown-bag lunch in Weinberg's capacious office
at which students or visitors gave short presentations.
He listened attentively and made comments and
suggestions. He was a nurturer of young physicists
and good group leader. \\

\noindent
For a few weeks, we worked on a research project together. It is so long ago we
cannot remember the topic. That was the
nearest we ever came to coauthor a paper.
We are saddened that,
with his passing, such coauthorship will never happen.
We recall that Steve regarded our calculations as competitions and, at that time,
he would give us a call at home, typically at 07:00 am in the morning,
to compare our separate progress. 

\bigskip

\noindent
\underline{A couple of memories from Austin:}\\

\noindent
(1). At one brown-bag lunch there were four talks, all on quite different
topics. At a faculty party that evening, somebody asked Weinberg
what the lunchtime talks had been about. What we noticed is that he could
recall many peculiarly verbatim
details and explain perspicuously
every talk so we believe he had at least a several-hours
echoic memory. Of course, everybody has an
echoic memory of several seconds otherwise discussion would be impossible. \\ 

\bigskip

\noindent
(2). For one lunch at the faculty club there were a dozen particle theorists at Weinberg's table, 
mostly dressed far too sloppily.
At a nearby table was the UT-Austin Board of Trustees, all elegantly dressed and
looking affluent.
We were sitting next to SW and, during a lull in the particle theory discussion, he looked over at the
other table and said\\
\noindent
"I sometimes wish I had decided to become rich, rather than famous".
We suspect Weinberg did not really believe that for one millisecond.

\section*{Acknowledgements}

\noindent
We thank the University of Salento for an affiliation. We thank S. Odintsov for 
advice on preparing this article which is based on a talk given December 18th, 2021
at the MIAMI2021 Conference.

\end{document}